\documentstyle[titlepage,12pt]{article}
\begin{document}
\renewcommand{\theequation}{\thesection.\arabic{equation}}
\title{New Classes of Exact Multi-String Solutions in Curved
Spacetimes}
\author{A.L. Larsen and N. S\'{a}nchez\\
\\
\\
Observatoire de Paris,
DEMIRM. Laboratoire Associ\'{e} au CNRS \\
\hspace*{-9mm}UA 336, Observatoire de Paris et
\'{E}cole Normale Sup\'{e}rieure. \\
\hspace*{-32mm}61, Avenue de l'Observatoire, 75014 Paris, France.}
\maketitle
\begin{abstract}
We find new classes of {\it exact} string solutions in a variety of curved
backgrounds. They include stationary and dynamical (open, closed, straight,
finitely and infinitely long) strings as well as {\it multi-string} solutions,
in terms of elliptic functions. The physical properties, string length,
energy and pressure are computed and analyzed. In anti de Sitter spacetime,
the solutions describe an {\it infinite}
number of infinitely long stationary strings
of equal energy but different pressures. In de Sitter spacetime, outside the
horizon, they describe infinitely many {\it dynamical} strings infalling
non-radially, scattering at the horizon and going back to spatial infinity
in different directions. For special values of
the constants of motion, there are families of solutions
with {\it selected finite}
numbers of different and independent strings. In black hole
spacetimes (without cosmological constant),
{\it no} multi-string solutions are found. In the Schwarzschild black hole,
inside the horizon, we find one
straight string infalling non-radially, with {\it indefinetely} growing size,
into the $r=0$ singularity. In the $2+1$ black hole anti de Sitter
background, the string stops at $r=0$ with {\it finite} length.
\end{abstract}
\section{Introduction and Results}
In this paper we continue our investigations on the {\it exact} string
dynamics in curved spacetimes, see for example Refs.[1-3].
In curved spacetimes, it is generally impossible to obtain the exact and
complete solution to the string equations of motion and constraints. These
are highly non-linear coupled partial differential equations, and are
generally not integrable. However, in some curved spacetimes like
gravitational wave backgrounds \cite{san4} and cosmic string
backgrounds \cite{san5}, the string dynamics turns out to be exactly solvable.
Moreover, the full string equations of motion and constraints have been
shown to be exactly integrable in $D$-dimensional de Sitter spacetime
\cite{san6}; they are equivalent to a generalized sinh-Gordon equation.
Several new properties like the multi-string solutions
\cite{mik1,mik2,san1} emerged.
On the other hand, in most spacetimes, quite general families of
exact solutions can be found by making an appropriate ansatz, which exploits
the symmetries of the underlying curved spacetime. In axially symmetric
spacetimes, a convenient ansatz corresponds to circular strings. Such an
ansatz effectively decouples the dependence on the spatial world-sheet
coordinate $\sigma,$ and the string equations of motion and constraints
reduce to non-linear coupled ordinary differential equations. They are
considerably simpler to handle than the original system, and they have indeed
been analyzed and solved in a number of interesting cases from
gravitation \cite{san2,all1},
cosmology \cite{san1,san3} and string theory \cite{san2,egu}. In
this paper we will make instead
an ansatz which effectively decouples the dependence on the temporal
world-sheet coordinate $\tau.$ This ansatz, which we call the "stationary
string ansatz" is dual to the "circular string ansatz" in the sense that it
corresponds to a formal interchange of the world-sheet coordinates
$(\tau,\sigma),$ as well as of the azimuthal angle $\phi$ and the
stationary time $t$ in the target space:
\begin{equation}
\tau\;\leftrightarrow\;\sigma,\;\;\;\;\;\;\;\;t\;\leftrightarrow\;\phi.
\end{equation}
The stationary string ansatz will describe stationary strings when $t$
(and $\tau$) are timelike, for instance in anti de Sitter spacetime
(in static coordinates) and outside the horizon of a Schwarzschild black hole.
On the other hand, if $t$ (and $\tau$) are spacelike, for instance inside the
horizon of a Schwarzschild black hole or outside the horizon of de Sitter
spacetime (in static coordinates), the stationary string ansatz will describe
dynamical propagating strings. Considering a static line element in the form:
\begin{equation}
ds^2=-a(r)dt^2+\frac{dr^2}{a(r)}+r^2(d\theta^2+\sin^2\theta d\phi^2),
\end{equation}
the stationary string ansatz reads explicitly:
\begin{equation}
t=\tau,\;\;\;\;\;r=r(\sigma),\;\;\;\;\;\phi=\phi(\sigma),\;\;\;\;\;\theta=
\pi/2.
\end{equation}
The string equations of motion and constraints reduce to two separated
first order ordinary differential equations:
\begin{equation}
\phi'=\frac{L}{r^2},\;\;\;\;\;\;r'^2+V(r)=0;\;\;\;\;
V(r)=-a(r)[a(r)-\frac{L^2}{r^2}],
\end{equation}
where $L$ is an integration constant.
The qualitative features of the possible string configurations can be read off
directly from the shape of the potential $V(r).$ Thereafter, the detailed
analysis of the quantitative features can be performed by explicitly solving
the (integrable) system of equations (1.4). The induced line element on the
world-sheet is given by:
\begin{equation}
ds^2=a(r)(-d\tau^2+d\sigma^2).
\end{equation}
Thus, if $a(r)$ is negative, the world-sheet coordinate $\tau$ becomes
spacelike while $\sigma$ becomes timelike and the stationary string
ansatz (1.3) describes a dynamical string. If $a(r)$ is positive, the ansatz
describes a stationary string.
\vskip 6pt
\hspace*{-6mm}In
this paper we solve explicitly Eqs.(1.4) and we analyze in detail the
solutions and their physical interpretation in Minkowski, de Sitter,
anti de Sitter, Schwarzschild and $2+1$ black hole anti de Sitter spacetimes.
In all these cases, the solutions are expressed in terms of elliptic
(or elementary) functions. We furthermore analyze the physical properties,
string length, energy and pressure of these solutions. We also give the
qualitative features of the solutions in Schwarzschild-de Sitter and
Schwarzschild-anti de Sitter spacetimes.

In Minkowski spacetime (M), the potential is given by:
\begin{equation}
V_{\mbox{M}}(r)=\frac{L^2}{r^2}-1,
\end{equation}
and the solution of Eqs.(1.4) describes one infinitely long straight string
with "impact-parameter" $L.$ The equation of state takes the well-known form:
\begin{equation}
dE=-dP_2=\mbox{const.},\;\;P_1=P_3=0.
\end{equation}
In anti de Sitter spacetime (AdS), the potential is given by:
\begin{equation}
V_{\mbox{AdS}}(r)=-(1+H^2r^2)[1+H^2r^2-\frac{L^2}{r^2}].
\end{equation}
The radial coordinate $r(\sigma)$ is periodic with finite period $T_\sigma,$
which is expressed in terms of a complete elliptic integral, Eq.(4.11). For
$\sigma\in[0,T_\sigma],$ the solution describes an infinitely long stationary
string in the wedge $\phi\in\;]0,\Delta\phi[\;,$ where:
\begin{equation}
\Delta\phi=2k\sqrt{\frac{1-2k^2}{1-k^2}}[\Pi(1-k^2,\;k)-K(k)]\;\in\;\;]0,\;\pi[
\end{equation}
The elliptic modulus $k$ parametrizes the solutions, $k\in\;]0,1/\sqrt{2}[\;.$
The azimuthal angle is generally not a periodic function of $\sigma,$
thus when the spacelike world-sheet coordinate $\sigma$ runs through the range
$]-\infty,+\infty[\;,$ the solution describes an {\it infinite} number
of infinitely long stationary open strings. The general solution is therefore
a multi-string solution. Until now multi-string solutions were only found
in de Sitter spacetime \cite{san1,mik1,mik2}. Our
results show that multi-string
solutions are a general feature of spacetimes with a cosmological
constant (positive or negative). The solution in anti de Sitter spacetime
describes a {\it finite} number of strings if the following relation holds:
\begin{equation}
N\Delta\phi=2\pi M.
\end{equation}
Here $N$ and $M$ are integers, determining the number of strings and the
winding in azimuthal angle, respectively, for the multi-string solution,
see Fig.2. The
equation of state for a full multi-string solution takes the form $(P_3=0)$:
\begin{equation}
dP_1=dP_2=-\frac{1}{2}dE,\;\;\;\;\;\mbox{for}\;\;r\rightarrow\infty
\end{equation}
corresponding to extremely unstable strings \cite{ven}.

In de Sitter spacetime (dS), the potential is given by:
\begin{equation}
V_{\mbox{dS}}(r)=-(1-H^2r^2)[1-H^2r^2-\frac{L^2}{r^2}].
\end{equation}
In this case we have to distinguish between solutions inside the horizon (where
$\tau$ is timelike) and solutions outside the horizon (where $\tau$ is
spacelike). Inside the horizon, the generic
solution describes one infinitely long open
stationary string winding around $r=0.$
For special values of the constants of motion, corresponding to a relation,
which formally takes the same form as Eq.(1.10), the solution describes a
closed string of finite length $l=N\pi/H.$ The integer $N$ in this case
determines the number of "leaves", see Fig.3. The energy is positive and
finite and grows with $N.$ The pressure turns out to vanish identically,
thus the equation of state corresponds to cold matter.
Outside the horizon, the world-sheet coordinate $\tau$ becomes spacelike
while $\sigma$ becomes timelike, thus we define:
\begin{equation}
\tilde{\tau}\equiv\sigma,\;\;\;\;\;\;\;\;\tilde{\sigma}\equiv\tau,
\end{equation}
and the string solution is expressed in hyperboloid coordinates, Eqs.(5.31).
The radial coordinate $r(\tilde{\tau})$ is periodic with a finite period
$T_{\tilde{\tau}},$ Eq.(5.29). For $\tilde{\tau}\in[0,T_{\tilde{\tau}}],$
the solution describes a straight string incoming non-radially
from spatial infinity, scattering at the horizon and escaping towards infinity
again, Fig.4. The string length is zero at the horizon and grows
indefinetely in the asymptotic regions.
As in the case of anti de Sitter spacetime,
the azimuthal angle is generally not a periodic function,
thus when the timelike
world-sheet coordinate $\tilde{\tau}$ runs through the range
$]-\infty,+\infty[\;,$ the solution describes an {\it infinite} number
of dynamical straight strings scattering at the horizon at different
angles. The general solution is therefore
a multi-string solution. In particular, a multi-string solution describing a
{\it finite} number of strings is obtained if a relation of the form (1.10) is
fulfilled. It turns out that the solution describes {\it at least}
three strings.
The energy and pressures of a full multi-string solution
are also computed. In the
asymptotic region they fulfill an equation of state corresponding to extremely
unstable strings, i.e. like Eq.(1.11).

In the Schwarzschild black hole background (S), the potential is given by:
\begin{equation}
V_{\mbox{S}}(r)=-(1-2m/r)[1-2m/r-\frac{L^2}{r^2}].
\end{equation}
{\it No} multi-string
solutions are found in this case.
Outside the horizon the solution of Eqs.(1.4) describes one infinitely long
stationary open string. This solution was derived in Ref.\cite{fro} and we
shall
not go into details here. In our notation the solution is given by
Eqs.(6.4)-(6.6). Inside the horizon, where $\tau$ becomes spacelike
while $\sigma$ becomes timelike, we make the redefinitions (1.13) and the
solution is expressed in terms of Kruskal coordinates, Eq.(6.16). The
solution describes one straight string infalling {\it non-radially} towards the
singularity. At the horizon, the string length is zero and it grows
{\it indefinitely} when the string approaches the spacetime
singularity.  Thereafter,
the solution can not be continued.

In the $2+1$ black hole anti de Sitter spacetime (BH-AdS) \cite{ban},
the potential is given by:
\begin{equation}
V_{\mbox{BH-AdS}}(r)=-(\frac{r^2}{l^2}-1)[\frac{r^2}{l^2}-1-\frac{L^2}{r^2}].
\end{equation}
Outside the horizon, the solutions "interpolate" between the solutions
found in anti de Sitter spacetime and outside the horizon of the
Schwarzschild black hole. The solutions thus describe infinitely long
stationary open strings. As in anti de Sitter spacetime, the general solution
is a {\it multi-string} describing {\it infinitely}
many strings. In particular, for certain
values of the constants of motion, corresponding to the condition of the form
(1.10), the solution describes a finite number of strings. In the simplest
version of the $2+1$ black hole anti de Sitter background ($M=1,\;J=0$),
it turns out that the solution describes {\it at least} seven
strings. Inside the
horizon, we make the redefinitions (1.13) and the solution is expressed in
terms of Kruskal-like coordinates, Eq.(7.21). The solution is
similar to the solution found inside the horizon of the
Schwarzschild black hole, but there is one important difference at $r=0.$
As in the Schwarzschild black hole background, the solution
describes one straight string infalling non-radially towards $r=0,$ and
beyond  this
point the solution can not be continued because of the global structure of the
spacetime. At the horizon the string size is zero and during the fall
towards $r=0,$ the string size {\it grows}
but stays {\it finite}. This should be
compared with the straight string inside the horizon of the
Schwarzschild black hole, where the string size grows indefinetely. The
physical reason for this difference is that the point $r=0$ is not a
strong curvature singularity in the $2+1$ black hole anti de Sitter
spacetime.

Finally we consider also the Schwarzschild de Sitter and
Schwarzschild anti de Sitter spacetimes. These spacetimes contain all
the features of the spacetimes already discussed: singularities, horizons,
positive or negative cosmological constants. All the various types of
string solutions found in the other spacetimes (open, closed, straight,
finitely and infinitely long, multi-strings) are therefore present in
the different regions of the Schwarzschild-de Sitter and
Schwarzschild-anti de Sitter spacetimes. The details are given in Section 8.

Throughout the paper we use sign conventions of Misner, Thorne and Wheeler
\cite{mis} and units in which, beside $G=1,\;c=1,$ the string
tension $(2\pi\alpha')^{-1}=1.$ A short summary of our results is presented
in Table I.
\section{General Formalism}
\setcounter{equation}{0}
In this section we derive the ordinary differential equations obtained from
the generic string equations of motion and constraints using a stationary
string ansatz. For simplicity we consider stationary strings embedded in
static spherically symmetric spacetimes. The results, however, can be easily
generalized to stationary axially symmetric spacetimes. Stationary strings
in stationary background spacetimes were first discussed from a
general point of view in Ref.\cite{fro}. It was shown that stationary string
configurations can be described by a geodesic equation in a properly
chosen "internal" space \cite{fro}. Using the same formalism, small
perturbations around the stationary strings have also been considered
\cite{fro2,all2}.
In the present paper we will use a different approach and we will find
with it new classes of solutions. For a
stationary string in a general static and spherically symmetric spacetime, we
derive an effective potential in the radial spacetime coordinate. The
potential provides  immediate information about the stationary string
configurations. To be more specific we consider the following
spacetime line element:
\begin{equation}
ds^2=-a(r)dt^2+\frac{dr^2}{a(r)}+r^2(d\theta^2+\sin^2\theta d\phi^2),
\end{equation}
which includes as special cases Minkowski spacetime, anti de Sitter spacetime,
de Sitter spacetime, Schwarzschild black hole spacetime and its
extensions including charge and cosmological constant. The string
equations of motion and constraints are:
\begin{eqnarray}
&\ddot{x}^\mu-x''^\mu+\Gamma^\mu_{\rho\sigma}(\dot{x}^\rho\dot{x}^\sigma-
x'^\rho x'^\sigma)=0,&\nonumber\\
&g_{\mu\nu}\dot{x}^\mu x'^\nu=g_{\mu\nu}(\dot{x}^\mu\dot{x}^\nu+x'^\mu x'^\nu)
=0,&\nonumber
\end{eqnarray}
where dot and prime stand for derivative with respect to $\tau$ and
$\sigma,$ respectively. For the metric defined by the line element (2.1),
they take the form:
\begin{eqnarray}
\ddot{t}\hspace*{-2mm}&-&\hspace*{-2mm}t''+\frac{a_{,r}}{a}(\dot{t}\dot{r}-
t'r')=0,\nonumber\\
\ddot{r}\hspace*{-2mm}&-&\hspace*{-2mm}r''-\frac{a_{,r}}{2a}(\dot{r}^2-r'^2)+
\frac{aa_{,r}}{2}(\dot{t}^2-t'^2)-ar(\dot{\theta}^2-\theta'^2)-ar\sin^2\theta
(\dot{\phi^2}-\phi'^2)=0,\nonumber\\
\ddot{\theta}\hspace*{-2mm}&-&\hspace*{-2mm}\theta''+\frac{2}{r}(\dot{\theta}
\dot{r}-\theta' r')-\sin\theta\cos\theta(\dot{\phi}^2-\phi'^2)=0,\nonumber\\
\ddot{\phi}\hspace*{-2mm}&-&\hspace*{-2mm}\phi''+\frac{2}{r}(\dot{\phi}
\dot{r}-\phi'
r')+2\cot\theta(\dot{\theta}\dot{\phi}-\theta'\phi')=0,\nonumber\\
\hspace*{-2mm}&-&\hspace*{-2mm}a\dot{t}t'+\frac{1}{a}\dot{r}r'+r^2\dot{\theta}
\theta'+r^2\sin^2\theta\dot{\phi}\phi'=0,\nonumber\\
\hspace*{-2mm}&-&\hspace*{-2mm}a(\dot{t}^2+t'^2)+\frac{1}{a}(\dot{r}^2+r'^2)
+r^2(\dot{\theta}^2+\theta'^2)+r^2\sin^2\theta(\dot{\phi}^2+\phi'^2)=0.
\end{eqnarray}
The stationary string ansatz, consistent with the
symmetries of the background, is taken to be:
\begin{equation}
t=\tau,\;\;\;\;\;r=r(\sigma),\;\;\;\;\;\phi=\phi(\sigma),\;\;\;\;\;\theta=
\pi/2,
\end{equation}
i.e. the string is in the equatorial plane and the two functions $r(\sigma),\;
\phi(\sigma)$ are to be determined by the equations of motion and constraints,
Eqs.(2.2). After inserting the ansatz, the equations of motion and constraints
are consistently reduced to two first order ordinary differential equations:
\begin{equation}
\phi'=\frac{L}{r^2},
\end{equation}
\begin{equation}
r'^2=a(r)[a(r)-\frac{L^2}{r^2}],
\end{equation}
where $L$ is an integration constant, which without loss of generality
can be taken to be non-negative. $r(\sigma)$ is obtained by inversion
of the integral:
\begin{equation}
\sigma-\sigma_0=\int_{r_0}^{r}\frac{dx}{\sqrt{a(x)[a(x)-(L/x)^2]}},
\end{equation}
after which $\phi(\sigma)$ is obtained by integrating Eq.(2.4). In all cases
under consideration in this paper, Eqs.(2.4)-(2.5) will be solved in terms
of elliptic or elementary functions. It is convenient to define an effective
potential $V(r)$ by:
\begin{equation}
V(r)=-a(r)[a(r)-\frac{L^2}{r^2}],
\end{equation}
such that the $r$-equation of motion takes the form:
\begin{equation}
r'^2+V(r)=0.
\end{equation}
With this definition the stationary string will be located at the $r$-axis
in a $(r,V(r))$ diagram. The possible string configurations can therefore
be read off from knowledge about the zeros of the potential. The exact
string shape can thereafter be obtained by solving Eqs.(2.4)-(2.5).
Notice that the circular string solutions of Eq.(2.8) must be considered
separately. They are determined by:
\begin{equation}
r=\mbox{const.}\equiv r_c,\;\;\;\;\;\;\phi=\frac{L}{r_c^2}\sigma,\;\;\;\;\;
V(r_c)=0,
\end{equation}
where $r_c\neq 0,\;L\neq 0,$ but they
are solutions to the original second order differential equations (2.2)
only provided:
\begin{eqnarray}
\frac{dV(r)}{dr}|_{r=r_c}=0.
\end{eqnarray}
We shall return to this point later for each of the curved backgrounds under
consideration.

Insertion of
the ansatz, using the results (2.4)-(2.5), in the line element (2.1), leads to:
\begin{equation}
ds^2=a(r)(-d\tau^2+d\sigma^2).
\end{equation}
The string length element $dl$ is then identified as:
\begin{equation}
dl=\sqrt{a(r(\sigma))}\;d\sigma,
\end{equation}
and the physical string length is obtained by integrating over the
appropriate range of $\sigma,$ which may be finite or infinite. For strings in
FRW-universes, it is interesting to also consider the string energy and
pressure that can be obtained from the spacetime energy-momentum tensor:
\begin{equation}
\sqrt{-g}T^{\mu\nu}=\int d\tau d\sigma (\dot{X}^\mu\dot{X}^\nu-
X'^\mu X'^\nu)\delta^{(4)}(X-X(\tau,\sigma)),
\end{equation}
by integrating over a volume that completely encloses the string.
The coordinates $X^\mu$ are here the comoving FRW-coordinates:
\begin{equation}
ds^2=-(dX^0)^2+a^2(X^0)\;\frac{dR^2+R^2(d\theta^2+
\sin^2\theta d\phi^2)}{(1+\frac{k}{4}R^2)^2},
\end{equation}
including as special cases Minkowski, de
Sitter and anti de Sitter
spacetimes:
\begin{eqnarray}
a(X^0) &=& 1,\;\;k=0 \quad{\rm for~Minkowski~ spacetime},\nonumber \\
a(X^0)&=&e^{HX^0},\;\;k=0 \quad{\rm for~de~ Sitter~ spacetime},\nonumber \\
a(X^0)&=&\cos(HX^0),\;\;k=-H^2
\quad{\rm for~anti~ de~ Sitter ~spacetime},\nonumber
\end{eqnarray}
which can all be brought into the static spherically symmetric form of
Eq.(2.1).

We close this section with the following interesting observation: the function
$a(r)$ introduced in the line element (2.1) is not necessarily non-negative.
In de Sitter and Schwarzschild spacetimes it changes sign at the horizon. Then
the world-sheet coordinate $\tau$ becomes spacelike while $\sigma$ becomes
timelike, see Eq.(2.11), and the string solution must be expressed in other
coordinates. In such cases the stationary string ansatz, Eq.(2.3), actually
describes dynamical propagating strings. We therefore reach the interesting
conclusion that the stationary string ansatz, Eq.(2.3), for different initial
conditions, describes both stationary equilibrium string
configurations as well as dynamical propagating strings, in different regions
of the background spacetime.

In the following sections we use the general formalism of this section to
describe and analyze stationary and dynamical strings in various spacetimes.
For completeness we first recall the results in flat Minkowski spacetime.
We then consider anti de Sitter and de Sitter spacetimes and thereafter turn
to the 2+1 and 3+1 dimensional black hole spacetimes with and
without cosmological constant.
\section{Minkowski Spacetime}
\setcounter{equation}{0}
In Minkowski spacetime the situation is considerably simple. The potential,
Eq.(2.7), takes the form (se Fig.1a):
\begin{equation}
V(r)=\frac{L^2}{r^2}-1,
\end{equation}
so that stationary strings are only possible in the region $r\geq L.$
Eqs.(2.4)-(2.5) are easily solved by:
\begin{equation}
r(\sigma)=\sqrt{\sigma^2+L^2},
\end{equation}
\begin{equation}
\phi(\sigma)=\arctan(\sigma/L),
\end{equation}
which for $\sigma\in\;]-\infty,\;+\infty[\;$ describes an infinitely long
straight string parallel to the $X^2$-axis with "impact-parameter" $L.$
Obviously, the string length, string energy and integrated string pressure
are all infinite. For infinitely long strings it is however more interesting
to consider the energy and pressure densities. From Eqs.(2.12)-(2.13) we
find:
\begin{equation}
\frac{dE}{dl}=\frac{d}{dl}\int d^3 X \sqrt{-g}T^{00}=1,
\end{equation}
\begin{equation}
\frac{dP_2}{dl}=\frac{d}{dl}\int d^3 X \sqrt{-g}T^{y}\;_{y}=-1,
\end{equation}
while $P_1=P_3=0.$ This is the well known result concerning stationary
strings in flat Minkowski spacetime. Finally, notice that the circular
string $r=L,\;\phi=\sigma/L,$ which solves the first order differential
equations (2.4)-(2.5), must be
excluded since it does not fulfill the original second order differential
equations of motion, Eq.(2.2), c.f. the remarks related to Eqs.(2.9)-(2.10).
\section{Anti de Sitter Spacetime}
\setcounter{equation}{0}
In anti de Sitter spacetime, which corresponds to $a(r)=1+H^2r^2,$ we find the
stationary string potential:
\begin{equation}
V(r)=-(1+H^2r^2)[1+H^2r^2-\frac{L^2}{r^2}],
\end{equation}
see Fig.1b. It follows that stationary strings can only be found for $r\geq
r_0:$
\begin{equation}
r_0=\frac{1}{H}\left( \frac{-1+\sqrt{1+4H^2L^2}}{2}\right)^{1/2}.
\end{equation}
The circular string configuration $r=r_0,\;\phi=L\sigma/r_0^2,$ which solves
the first order differential
equations (2.4)-(2.5), is
excluded since it does not fulfill the original second order differential
equations of motion, Eq.(2.2), c.f. the remarks related to Eqs.(2.9)-(2.10).
This is similar to the situation in
Minkowski spacetime. It turns out then, that all stationary strings in
anti de Sitter spacetime are infinitely long open strings. For $L=0\;(r_0=0)$
we have
the straight string on $r=0:$
\begin{equation}
r=\frac{1}{H}\tan(H\sigma),\;\;\;\;\;\;\phi=\mbox{const.}
\end{equation}
For $L\neq 0,$ Eq.(2.5) is solved by a Weierstrass elliptic function:
\begin{equation}
H^2r^2(\sigma)=\frac{1}{H^2}\wp(\sigma-\sigma_0;\;g_2,\;g_3)-\frac{2}{3},
\end{equation}
with invariants:
\begin{equation}
g_2=4H^4(\frac{1}{3}+H^2L^2),\;\;\;\;\;g_3=\frac{4}{3}H^6(\frac{2}{9}+H^2L^2),
\end{equation}
discriminant:
\begin{equation}
\Delta=16H^{16}L^4(1+4H^2L^2)
\end{equation}
and roots:
\begin{equation}
e_1=\frac{H^2}{6}[1+3\sqrt{1+4H^2L^2}\;]>e_2=-\frac{H^2}{3}>
e_3=\frac{H^2}{6}[1-3\sqrt{1+4H^2L^2}\;].
\end{equation}
The integration constant $\sigma_0$ must be carefully chosen in order to
obtain a real
solution for real $\sigma.$ In the present case it turns out that $\sigma_0$
must be real and it is convenient to take $\sigma_0=0.$ Then, the solution
(4.4) can be written in terms of a Jacobi elliptic function:
\begin{equation}
H^2r^2(\sigma)=\nu^2-\mu^2+\mu^2\;\mbox{ns}^2[\mu H\sigma,\;k],
\end{equation}
where we introduced the more compact notation:
\begin{equation}
\mu^2\equiv\sqrt{1+4H^2L^2},\;\;\;\;\;
\nu^2\equiv\frac{1}{2}(-1+\mu^2\;),
\;\;\;\;\;k\equiv\frac{\nu}{\mu}.
\end{equation}
It follows that the elliptic modulus $k\in\;]0,\;1/\sqrt{2}[\;,$ where
$k\rightarrow 0$ corresponds
to $L\rightarrow 0,$ while $k\rightarrow 1/\sqrt{2}$ corresponds to
$L\rightarrow\infty.$
By integrating Eq.(2.4) we find:
\begin{equation}
\phi(\sigma)=\frac{-k}{\sqrt{1-k^2}}\left\{H\sigma-\sqrt{1-2k^2}\;\Pi(1-k^2,\;
\mu H\sigma,\;k)\right\},
\end{equation}
where $\Pi$ is the elliptic integral of the third kind.
Eqs.(4.8), (4.10) provide the complete solution for stationary strings in
anti de Sitter spacetime. The radial coordinate, Eq.(4.8), is periodic
with period:
\begin{equation}
T_\sigma=\frac{2K(k)}{\mu H},
\end{equation}
where $K(k)$ is the complete elliptic integral of the first kind. For
$\sigma\in[0,\;T_\sigma],\;r$ goes from infinity towards $r=r_0$ and
back towards infinity. In the same range of $\sigma,$ the Azimuthal
angle $\phi$ goes from $\phi=0$ to $\phi=\Delta\phi:$
\begin{equation}
\Delta\phi=2k\sqrt{\frac{1-2k^2}{1-k^2}}[\Pi(1-k^2,\;k)-K(k)]\;\in\;\;]0,\;\pi[
\end{equation}
that is to say,
$\Delta\phi$ is the angle between the two "arms" of the stationary string.
Two important remarks are now in order: First, the stationary string extends
from spatial infinity to spatial infinity for a {\it finite} range of the
world-sheet coordinate $\sigma.$ Secondly, the Azimuthal angle is generically
{\it not} a periodic function of $\sigma$ with period $T_\sigma$ (or an integer
multiple of $T_\sigma$),
not even modulo $2\pi.$ These two statements together imply that the
solution described by Eqs.(4.8), (4.10) is actually a {\it multi-string}
solution in the sense that one single world-sheet, determined by one set
of initial conditions, describes a finite or even an infinite number of
different and independent strings.
Multi-string solutions were first found in de
Sitter spacetime for {\it non-stationary} strings.
\cite{mik1,mik2,san1}. In that case, it was found that
strings can contract from infinite
size to a minimal size and back towards infinite size for a {\it finite} range
of the timelike world-sheet coordinate $\tau.$ The process, on the other hand,
takes infinite physical time, such that when the world-sheet time $\tau$
runs from $-\infty$ to $+\infty,$ the world-sheet describes {\it infinitely}
(for degenerate cases: finitely) many strings. The physical time was not
periodic so the solution really described different and independent strings;
not simply
infinitely many (or finitely many)
copies of the same string. Here, in anti de Sitter
spacetime the situation is somewhat similar, but with the roles of $\tau$
and $\sigma$ as well as $t$ and $\phi$ interchanged: For
$\sigma\in[0,\;T_\sigma]$ the solution describes one infinitely long string
in the wedge $\phi\in[0,\;\Delta\phi],$ for $\sigma\in[T_\sigma,\;2T_\sigma]$
it describes another infinitely long string in the wedge
$\phi\in[\Delta\phi,\; 2\Delta\phi]$ etc. In the general case, the solution,
Eqs.(4.8), (4.10), describes infinitely many strings. However,
for certain values
of the "impact-parameter" $L,$ or alternatively of the elliptic modulus $k,$
the Azimuthal angle becomes periodic modulo $2\pi$ and then the solution
describes only a finite number of strings. Clearly, this situation appears
provided:
\begin{equation}
N\Delta\phi=2\pi M,
\end{equation}
where $N$ and $M$ are integers. From the exact expression of $\Delta\phi,$
Eq.(4.12), it follows that $M/N\in\;]0,1/2[,\;$ and the solutions are then
conveniently parametrized in terms of $(N,M).$ The simplest examples
are $(N,M)=(3,1),\;(N,M)=(4,1)$ etc., see Fig.2.
The integer $N$ gives the number
of strings in the multi-string solution while the integer $M$ is a
"winding-number" of
the Azimuthal angle. It should be stressed also that the general
solution is a multi-string solution but it can of course be truncated,
in particular,
to describe only one string by considering an appropriate range of
the world-sheet coordinate $\sigma.$

We now consider the physical properties of the stationary strings found above.
Each string has infinite length; the string-length element, Eq.(2.12), being:
\begin{equation}
\frac{dl}{d\sigma}=
\sqrt{1+H^2r^2(\sigma)}=\frac{\mid ds[\mu H\sigma,\;k]\mid}{\sqrt{1-2k^2}},
\end{equation}
which diverges for $\sigma=0,\;\pm T_\sigma,\;\pm 2T_\sigma,...,$ i.e. at
spatial infinity. Next we consider the energy and pressure densities. The
energy as a function of the cosmic time is obtained by integrating $T^{00}:$
\begin{equation}
E(X^0)=\int d^3 X\sqrt{-g}\;T^{00},
\end{equation}
where $T^{\mu\nu}$ is given by Eq.(2.13), and the cosmic time $X^0$ is given
in terms of static coordinates by:
\begin{equation}
HX^0(\tau,\sigma)=
\pm\arccos\sqrt{(1+H^2r^2(\sigma))\cos^2(H\tau)-H^2r^2(\sigma)}.
\end{equation}
For a full $(N,M)$ multi-string solution we find after some algebra,
the following integral expression:
\begin{equation}
E(X^0)=2N\int_0^{T_\sigma/2}
d\sigma\frac{\mid\cos(HX^0)\mid[(1+H^2r^2(\sigma))^2+
H^2L^2\tan^2(HX^0)]}{(1+H^2r^2(\sigma))\sqrt{H^2r^2(\sigma)+\cos^2(HX^0)}}.
\end{equation}
Not surprisingly, the total energy is infinite. Using also Eq.(4.14), we
find the energy density:
\begin{equation}
\frac{dE(X^0)}{dl}=2N\;\frac{\mid\cos(HX^0)\mid[(1+H^2r^2(\sigma))^2+
H^2L^2\tan^2(HX^0)]}{(1+H^2r^2(\sigma))^{3/2}
\sqrt{H^2r^2(\sigma)+\cos^2(HX^0)}},
\end{equation}
with the asymptotic value:
\begin{equation}
\frac{dE(X^0)}{dl}=2N\;\mid\cos(HX^0)\mid,\;\;\;\mbox{for}\;\;r\rightarrow\infty
\end{equation}
Notice that the right hand side of Eq.(4.18)
can be expressed in terms of the physical
length $l$ (and $X^0$), by integrating Eq.(4.14), and that the asymptotic
region $l\rightarrow\infty$ corresponds to $r\rightarrow\infty.$
Each individual string in the multi-string solution gives the same
contribution to the energy density but different
contributions to the pressure densities in the different directions. However,
if we consider the full $(N,M)$ multi-string solution,
we find $P_3=0,\;P_1=P_2\equiv P:$
\begin{eqnarray}
P(X^0)\hspace{-2mm}&=&\hspace{-2mm}
\int d^3X\sqrt{-g}\;T^1\;_1=\int d^3X\sqrt{-g}\;T^2\;_2\\
\hspace{-2mm}&=&\hspace{-2mm}N\int d\tau\int_0^{T_\sigma/2}d\sigma
\frac{\cos^2(HX^0)[\dot{R}^2-R'^2-R^2\phi'^2]}{(1-H^2R^2/4)^2}
\delta(X^0-X^0(\tau,\sigma)),\nonumber
\end{eqnarray}
where $X^1=R\cos\phi,\;X^2=R\sin\phi$ (in the equatorial plane) and $R$
is expressed in terms of static coordinates by:
\begin{equation}
HR(\tau,\sigma)=\frac{2}{Hr(\sigma)}[\sqrt{1+H^2r^2(\sigma)}\;\cos(H\tau)-
\sqrt{(1+H^2r^2(\sigma))\cos^2(H\tau)-H^2r^2(\sigma)}\;],
\end{equation}
see Eqs.(2.1), (2.14). After some algebra we find:
\begin{equation}
P(X^0)=-N\int_0^{T_\sigma/2}
d\sigma\frac{\mid\cos(HX^0)\mid[(1+H^2r^2(\sigma))^2-
H^2L^2\tan^2(HX^0)]}{(1+H^2r^2(\sigma))\sqrt{H^2r^2(\sigma)+\cos^2(HX^0)}},
\end{equation}
which is infinite. The pressure density is finite:
\begin{equation}
\frac{dP(X^0)}{dl}=-N\;\frac{\mid\cos(HX^0)\mid[(1+H^2r^2(\sigma))^2-
H^2L^2\tan^2(HX^0)]}{(1+H^2r^2(\sigma))^{3/2}
\sqrt{H^2r^2(\sigma)+\cos^2(HX^0)}},
\end{equation}
with the asymptotic value:
\begin{equation}
\frac{dP(X^0)}{dl}=-N\;\mid\cos(HX^0)\mid,\;\;\;\mbox{for}\;\;r\rightarrow\infty
\end{equation}
Comparing with Eq.(4.19), the energy and pressure densities in the
asymptotic region fulfill:
\begin{equation}
dP(X^0)=-\frac{1}{2}dE(X^0),\;\;\;\mbox{for}\;\;r\rightarrow\infty
\end{equation}
This type of "equation of state" is quite
typical for strings in 2+1 dimensional FRW-universes [3,11,17], so
it is not
surprising that we recover it here for stationary strings in the equatorial
plane of 3+1 dimensional anti de Sitter spacetime.
Generally, for arbitrary $r(\sigma),$ there is no simple expression for the
equation of state and the pressure density can take both positive and negative
values depending on $X^0$ and $r(\sigma).$ An exception is the case when $L=0$
where the string solution (4.3) describes a straight string. If the string is
oriented along the $X^2$-axis we find:
\begin{equation}
E(X^0)=-P_2(X^0)=2\int_0^{\pi/2}d\sigma\;\frac{|\cos(HX^0)|(1+H^2r^2(\sigma))}
{\sqrt{H^2r^2(\sigma)+\cos^2(HX^0)}},
\end{equation}
while $P_1=P_3=0.$ The integrated energy and pressure are infinite but
the densities fulfill the same equation of state as the straight string in
flat Minkowski spacetime, compare with Eqs.(3.4)-(3.5).
\section{De Sitter Spacetime}
\setcounter{equation}{0}
We now come to the cosmologically interesting case of de Sitter spacetime,
corresponding to $a(r)=1-H^2r^2$ in Eq.(2.1).  The stationary string
potential in this case is:
\begin{equation}
V(r)=-(1-H^2r^2)[1-H^2r^2-\frac{L^2}{r^2}],
\end{equation}
see Fig.1c. The potential vanishes at the horizon $r=1/H$ and for $r=r_{0\pm}:$
\begin{equation}
r_{0\pm}=\frac{1}{H}\left(\frac{1\pm\sqrt{1-4H^2L^2}}{2}\right)^{1/2},
\end{equation}
and we must consider separately the two cases: $HL>1/2$ and $HL<1/2.$ The first
order differential equations (2.4)-(2.5) are solved by $r=r_{0\pm},\;\;
\phi=L\sigma/r^2_{0\pm}$ as well as by $r=1/H,\;\;\phi=H^2L\sigma,$
but these three solutions must be excluded (c.f. Eqs.(2.9)-(2.10)) since
they do not fulfill the original second order differential equations (2.2)
except in the case $HL=1/2,$ where one of them survives:
\begin{equation}
r=\frac{1}{\sqrt{2}H},\;\;\;\;\;\phi(\sigma)=H\sigma.
\end{equation}
This is the stationary circular string configuration in de Sitter spacetime
already discussed in Refs.\cite{san1,mik1,mik2,fro2}. Another
"degenerate" case appears for $L=0$
where we find the straight string configuration:
\begin{equation}
r(\sigma)=\frac{1}{H}\tanh(H\sigma),\;\;\;\;\;\phi=\mbox{const.}
\end{equation}
Let us now consider the general case $L\neq 0,\;\;HL\neq 1/2.$
\vskip 6pt
\hspace*{-6mm}{\bf I. $0<HL<1/2:$}\\
In this case the potential has two different real zeros inside the horizon
besides the zero at the horizon $r=1/H.$
It follows from the potential, Fig.1c, that solutions will exist both inside
the horizon and outside the horizon, but they can never actually cross the
horizon. The solutions outside the horizon must be expressed in terms of
comoving coordinates or hyperboloid coordinates (say), since the static
coordinates, Eq.(2.1), are only appropriate to cover
the region of de Sitter spacetime inside
the horizon; we will return to that question later.
Eq.(2.5) is solved by a Weierstrass elliptic function (compare
with Eqs.(4.4)-(4.7)):
\begin{equation}
H^2r^2(\sigma)=\frac{1}{H^2}\wp(\sigma-\sigma_0;\;g_2,\;g_3)+\frac{2}{3},
\end{equation}
with invariants:
\begin{equation}
g_2=4H^4(\frac{1}{3}-H^2L^2),\;\;\;\;\;g_3=\frac{4}{3}H^6(-\frac{2}{9}+H^2L^2),
\end{equation}
discriminant:
\begin{equation}
\Delta=16H^{16}L^4(1-4H^2L^2)
\end{equation}
and roots:
\begin{equation}
e_1=\frac{H^2}{3}>e_2=\frac{H^2}{6}[-1+3\sqrt{1-4H^2L^2}\;]>
e_3=\frac{H^2}{6}[-1-3\sqrt{1-4H^2L^2}\;].
\end{equation}
The solution, Eq.(5.5)-(5.8), was originally derived in Ref.\cite{fro2}, but it
was only analyzed in the two "degenerate" cases corresponding to the solutions
given by Eqs.(5.3)-(5.4). Here we shall analyze the general solution. We
can write Eq.(5.5) in terms of a Jacobi elliptic function:
\begin{equation}
H^2r^2(\sigma)=\nu^2+\mu^2\;\mbox{ns}^2[\mu H(\sigma-\sigma_0),\;k],
\end{equation}
where we introduced the more compact notation (compare with Eqs.(4.8)-(4.9)):
\begin{equation}
\mu^2\equiv\frac{1}{2}(1+\sqrt{1-4H^2L^2}),\;\;\;\;\;
\nu^2\equiv\frac{1}{2}(1-\sqrt{1-4H^2L^2}),
\;\;\;\;\;k\equiv\frac{\sqrt{\mu^2-\nu^2}}{\mu}.
\end{equation}
It follows that the elliptic modulus
$k\in\;]0,\;1[\;,$ where $k\rightarrow 0$ corresponds
to $HL\rightarrow 1/2,$ while $k\rightarrow 1$ corresponds to
$L\rightarrow 0.$ We still have not fixed the integration constant $\sigma_0.$
It turns out that two qualitatively different families of real solutions appear
depending on the choice of $\sigma_0.$ For $\sigma_0=0$ we find:
\begin{equation}
H^2r_+^2(\sigma)=\nu^2+\mu^2\;\mbox{ns}^2[\mu H\sigma,\;k],
\end{equation}
while for $\sigma_0=\omega'=iK'(k)/(\mu H):$
\begin{equation}
H^2r_-^2(\sigma)=\nu^2+(\mu^2-\nu^2)\;\mbox{sn}^2[\mu H\sigma,\;k].
\end{equation}
Notice that $H^2r_+^2(\sigma)\geq 1,$ thus this solution is always outside the
horizon and it must be expressed in a different coordinate system.
For the $r_-$-solution we find that $\mu^2\geq H^2r_-^2(\sigma)\geq \nu^2,$
thus this solution is always inside the horizon, in fact, it oscillates
(in the spacelike world-sheet coordinate $\sigma$) between $\nu^2>0$ and
$\mu^2<1.$

Let us first consider the solution $r_-$ in a little more detail.
The corresponding Azimuthal angle is obtained from Eq.(2.4):
\begin{equation}
\phi_-(\sigma)=\sqrt{\frac{2-k^2}{1-k^2}}\Pi(\frac{-k^2}{1-k^2},\;\mu H\sigma,
\;k),
\end{equation}
where $\Pi$ is the elliptic integral of the third kind. The radial
coordinate, Eq.(5.12), is periodic with a period $T_\sigma$ as
formally given by Eq.(4.11),
but with $\mu$ and $k$ given by Eq.(5.10). For $\sigma\in[0,\;T_\sigma],$
$\;r_-$ increases from $r_-=\nu/H$ to $r_-=\mu/H$ and then decreases back to
$r_-=\nu/H.$ In the same range of $\sigma,$ the Azimuthal angle increases from
$\phi_-=0$ to $\phi_-=\Delta\phi_-:$
\begin{equation}
\Delta\phi_-=2\sqrt{\frac{2-k^2}{1-k^2}}\Pi(\frac{-k^2}{1-k^2},\;k)\;\in\;\;
]\pi,\;\sqrt{2}\pi[
\end{equation}
Generally the solution described by Eqs.(5.12)-(5.13) represents an
{\it infinitely long open} string
winding around $r=0$ between $r=\nu/H$ and $r=\mu/H.$ However,
in the special case where the Azimuthal angle is periodic modulo $2\pi$
with a period which is an integer multiple of
$T_\sigma,$ it describes a closed string of {\it finite length}.
This closed string condition takes the form:
\begin{equation}
N\Delta\phi_-=2\pi M,
\end{equation}
which is similar to Eq.(4.13) but with $\Delta\phi$
replaced by $\Delta\phi_-.$ The closed strings inside the
horizon of de Sitter spacetime are then parametrized in terms of the two
integers $(N,M),$ and we find from Eq.(5.14):
\begin{equation}
M/N\in\;]1/2,\;1/\sqrt{2}[.
\end{equation}
The simplest examples are $(N,M)=(3,2),\;
(N,M)=(5,3)\;$ etc., see Fig.3. The integer $N$ gives the number of "leaves"
(see Fig.3) of the solution, while
the integer $M$ is a winding number of the Azimuthal angle.
At this point it is interesting also to determine which types of stationary
closed string solutions are {\it not} allowed because of the conditions,
Eqs.(5.15)-(5.16). It is for instance impossible to have a solution with 4
"leaves" $(N=4),$ while 3 "leaves" $(N=3)$ or 5 "leaves" $(N=5)$
are perfectly allowed. Notice also that the closed circular string, Eq.(5.3),
in a pure mathematical sense, corresponds to $(N,M)=(\sqrt{2},1).$

Using Eqs.(2.12)-(2.13) we can calculate the physical length, energy and
pressure of the $(N,M)$ strings described above. The string length is given by:
\begin{equation}
l_-=\int_0^{2NK(k)/(\mu H)}\hspace*{-4mm}
d\sigma\;\sqrt{1-H^2r_-^2(\sigma)}=N\pi/H.
\end{equation}
This result holds for the circular string also (taking formally $N=\sqrt{2},$
as discussed above). The string energy is given by an expression of the form
(4.15) where:
\begin{equation}
HX^0(\tau,\sigma)=\frac{1}{2}\log(1-H^2r_-^2(\sigma))+H\tau.
\end{equation}
Using Eq.(2.13), the integrals are easily evaluated and we find:
\begin{equation}
E_-(X^0)=E_-=\frac{4N}{H}\;\frac{E(k)}{\sqrt{2-k^2}},
\end{equation}
where $E(k)$ is the complete elliptic integral of the second kind.
It should be stressed that $N$ and $k$ are not independent quantities. For
each $(N,M)$ string, the elliptic modulus must be calculated by solving
Eq.(5.15). Numerically we find that the energy is an
increasing function of $N,$ i.e. the energy grows with the number of "leaves".
Notice that the result, Eq.(5.19), holds also for the
circular string ($N=\sqrt{2}$), in agreement with \cite{san1} (in units
where $(2\pi\alpha')^{-1}=1).$ Let us finally calculate the pressure of a
generic
$(N,M)$ string. From the symmetries of the problem, it follows that $P_3=0,\;\;
P_1=P_2\equiv P_-,$ where:
\begin{eqnarray}
P_-(X^0)\hspace*{-2mm}&=&\hspace*{-2mm}\int d^3X\sqrt{-g}\;T^1\;_1=
\int d^3X\sqrt{-g}\;T^2\;_2\\
\hspace*{-2mm}&=&\hspace*{-2mm}e^{2HX^0}\int d\tau\int_0^{2NK(k)/(\mu H)}
d\sigma[\dot{X^1}\dot{X^1}-X'^1 X'^1]\delta(X^0-X^0(\tau,\sigma)),\nonumber
\end{eqnarray}
and where:
\begin{equation}
X^1(\tau,\sigma)=\frac{r_-(\sigma)\cos\phi_-(\sigma)}
{\sqrt{1-H^2r^2_-(\sigma)}}e^{-H\tau},
\end{equation}
see Eqs.(2.1), (2.14). After some algebra the integral is reduced to:
\begin{equation}
P_-(X^0)=P_-=N\int_{\nu/H}^{\mu/H} dr\frac{(H^2r^2-1)^2+H^2L^2}
{(1-H^2r^2)^{3/2}\sqrt{1-H^2r^2-(L/r)^2}}.
\end{equation}
This integral can be evaluated in terms of complete elliptic integrals:
\begin{equation}
P_-=\frac{N}{H\sqrt{2-k^2}}[(1-k^2)\Pi(k^2,k)-E(k)],
\end{equation}
which
vanishes identically
by an identity between elliptic integrals (Ref.\cite{abr}, formula 17.7.24):
\begin{eqnarray}
\Pi(k^2,\arcsin(\mbox{sn}[u,k]),k)=
\frac{E[\arcsin(\mbox{sn}[u,k]),k]}{1-k^2}-\frac{k^2}{1-k^2}\;
\frac{\mbox{sn}[u,k]\sqrt{1-\mbox{sn}^2[u,k]}}{\sqrt{1-k^2\mbox{sn}^2[u,k]}}.
\nonumber
\end{eqnarray}
For $u=K(k)$ (in our notation), this is exactly the combination of elliptic
integrals appearing in Eq.(5.23).
The closed stationary strings inside the horizon in de Sitter spacetime thus
fulfill an equation of state of the cold matter type. A similar result was
found recently \cite{san3} for oscillating circular strings in de Sitter
spacetime.

We now consider in more detail the solution $r_+(\sigma),$ Eq.(5.11).
Although obtained from the stationary string ansatz, Eq.(2.3), this solution
is actually not stationary since it is always outside the horizon where
$\tau$ is spacelike while $\sigma$ is timelike, as follows from Eq.(2.11) when
$a(r)=1-H^2r^2.$ Therefore, we define:
\begin{equation}
\tilde{\sigma}\equiv\tau,\;\;\;\;\;\tilde{\tau}\equiv\sigma,
\end{equation}
and identify the physical string length element outside the horizon as:
\begin{equation}
dl=\sqrt{H^2r^2-1}\;d\tilde{\sigma},
\end{equation}
which vanishes at the horizon and diverges at infinity. The actual string
length is obtained by integrating over the range of $\tilde{\sigma},$
which is now a finite range,
for instance $\tilde{\sigma}\in[0,\;\pi].$ Since the
radial coordinate now depends on the timelike world-sheet coordinate
$\tilde{\tau},$ the string length is then simply:
\begin{equation}
l_+(\tilde{\tau})=\pi\sqrt{H^2r_+^2(\tilde{\tau})-1}.
\end{equation}
In our further analysis we shall not need to specify the exact range of
$\tilde{\sigma}.$
The solution (5.11) is written as:
\begin{equation}
H^2r_+^2(\tilde{\tau})=\nu^2+\mu^2\;\mbox{ns}^2[\mu H\tilde{\tau},\;k].
\end{equation}
The corresponding Azimuthal angle is obtained from Eqs.(2.4) and (5.27):
\begin{equation}
\phi_+(\tilde{\tau})=\frac{1}{\sqrt{1-k^2}}\left\{ H\tilde{\tau}-
\sqrt{2-k^2}\;\Pi(-(1-k^2),\;\mu H\tilde{\tau},\;k)\right\},
\end{equation}
thus it
also depends on the timelike world-sheet coordinate $\tilde{\tau},$
only. The radial coordinate, Eq.(5.27), oscillates between the horizon and
infinity with period:
\begin{equation}
T_{\tilde{\tau}}=\frac{2K(k)}{\mu H}.
\end{equation}
For $\tilde{\tau}\in[0,\;T_{\tilde{\tau}}],$ $r$ goes from infinity towards
the horizon and back towards infinity. In the same $\tilde{\tau}$-range, the
string size, Eq.(5.26), contracts from infinite size to zero size
and then expands
back to infinite size.
The Azimuthal angle increases from $\phi_+=0$ to $\phi_+=\Delta\phi_+:$
\begin{equation}
\Delta\phi_+=2\sqrt{\frac{2-k^2}{1-k^2}}[K(k)-\Pi(-(1-k^2),\;k)]\;\in\;\;
]0,\;\pi(\sqrt{2}-1)[
\end{equation}
In order to better obtain
the physical interpretation of this solution, it must be expressed in a
coordinate system which covers the complete de Sitter manifold. In
hyperboloid coordinates, the solution takes the form:
\begin{eqnarray}
q^0\hspace*{-2mm}&=&\hspace*{-2mm}q^0(\tilde{\tau},\tilde{\sigma})=
\sqrt{H^2r_+^2(\tilde{\tau})-1}\cosh(H\tilde{\sigma}),\nonumber\\
q^1\hspace*{-2mm}&=&\hspace*{-2mm}q^1(\tilde{\tau},\tilde{\sigma})=
\sqrt{H^2r_+^2(\tilde{\tau})-1}\sinh(H\tilde{\sigma}),\nonumber\\
q^2\hspace*{-2mm}&=&\hspace*{-2mm}q^2(\tilde{\tau})=
Hr_+(\tilde{\tau})\cos\phi_+(\tilde{\tau}),\nonumber\\
q^3\hspace*{-2mm}&=&\hspace*{-2mm}q^3(\tilde{\tau})=
Hr_+(\tilde{\tau})\sin\phi_+(\tilde{\tau}),
\end{eqnarray}
and we have dropped $q^4,$ which is identically zero. As usual, we should
take another copy of $(q^0,q^1)$ to cover the full de Sitter manifold;
the coordinates, Eq.(5.31), only describe the expanding de Sitter universe
($q^0\geq 0$). Notice that the spatial coordinates $q^{i}$
depend on the spacelike
world-sheet coordinate $\tilde{\sigma}$ through $q^1$ only, and that
all $\tilde{\sigma}$-dependence disappears at the horizon. The solution
$r_+$ thus describes a straight string which at $q^0=-\infty$ starts at
spatial infinity with infinite length. As the de Sitter universe contracts,
the straight string falls non-radially towards the horizon while it contracts
and eventually becomes a point at the horizon, when the de Sitter universe
takes
its minimal size (for $q^0=0$). As
the de Sitter universe expands, the straight string also expands
and travels non-radially away towards infinity, where its length grows
indefinitely. Since the string travels from spatial infinity towards the
horizon and back towards spatial infinity for $q^0\in\;]-\infty,\;+\infty[\;,$
but for a {\it finite} range of the world-sheet coordinate $\tilde{\tau}$
(say $\tilde{\tau}\in[0,\;T_{\tilde{\tau}}]$), the
full solution is actually a multi-string. In the general case, the solution
describes infinitely many straight strings incoming from different
directions, "scattering" at the horizon and then going out again in
different directions towards
infinity.
An exceptional case is provided by the limit $L=0,$ where the solution
describes only one string falling radially in from infinity and going radially
out again. In all other cases, the solution describes finitely or infinitely
many strings. Following the analysis of the stationary multi-strings in anti
de Sitter spacetime, as done in
Section 4, we find the following condition for the
solution to describe only finitely many strings in de Sitter spacetime:
\begin{equation}
N\Delta\phi_+=2\pi M,
\end{equation}
where $N$ and $M$ are again integers. It follows that
$M/N\in\;]0,\;1/\sqrt{2}-1/2[\;$ and the simplest examples are then
$(N,M)=(5,1),\;\;(N,M)=(6,1)$ etc., see Fig.4.
Again, the integer $N$ describes the
number of strings in the multi-string solution while the integer $M$ is
a winding-number of the Azimuthal angle.

In the limit $HL=1/2,$ the
solution, Eqs.(5.27)-(5.28), reduces to elementary functions:
\begin{equation}
H^2r_+^2(\tilde{\tau})=1+\frac{1}{2}\cot^2\frac{H\tilde{\tau}}{\sqrt{2}},
\end{equation}
\begin{equation}
\phi_+(\tilde{\tau})=H\tilde{\tau}-
\arctan(\sqrt{2}\tan\frac{H\tilde{\tau}}{\sqrt{2}}).
\end{equation}
For this solution we have $\Delta\phi_+=\pi(\sqrt{2}-1),$ and it describes
infinitely many strings.

We already considered the physical length of the string solutions,
Eqs.(5.27)-(5.28). Let us now consider the energy and pressure.
The string energy is given by an expression of the form
(4.15) where:
\begin{equation}
HX^0(\tilde{\tau},\tilde{\sigma})=
\frac{1}{2}\log(H^2r_+^2(\tilde{\tau})-1)+H\tilde{\sigma}.
\end{equation}
For a full $(N,M)$ multi-string we find using Eq.(2.13):
\begin{equation}
E_+(X^0)=2N\int d\tilde{\sigma}\frac{e^{3H(X^0-\tilde{\sigma})}
+H^2L^2e^{-H(X^0-\tilde{\sigma})}}{\sqrt{e^{4H(X^0-\tilde{\sigma})}+
e^{2H(X^0-\tilde{\sigma})}+H^2L^2}}.
\end{equation}
By the substitution $z=\mbox{Exp}{(H\tilde{\sigma})},$ the integral can be
evaluated in terms of elliptic integrals but we shall not give the
explicit expression here. Asymptotically, when the string travels towards
spatial infinity, we find from Eqs.(5.25), (5.36) the following
expression for the energy density:
\begin{equation}
\frac{dE_+(X^0)}{dl_+}=2N,\;\;\;\mbox{for}\;\;X^0\rightarrow\infty
\end{equation}
It follows that the energy diverges in this limit, since the physical length
diverges. Concerning the pressures, we find that $P_3=0,\;P_1=P_2\equiv P_+:$
\begin{eqnarray}
P_+(X^0)\hspace*{-2mm}&=&\hspace*{-2mm}\int d^3X\sqrt{-g}\;T^1\;_1=
\int d^3X\sqrt{-g}\;T^2\;_2\\
\hspace*{-2mm}&=&\hspace*{-2mm}e^{2HX^0}\int \hspace*{-2mm}\;d\tilde{\sigma}
\int_0^{2NK(k)/(\mu H)}
d\tilde{\tau}[(\frac{dX^1}{d\tilde{\tau}})^2
-(\frac{dX^1}{d\tilde{\sigma}})^2]
\delta(X^0-X^0(\tilde{\tau},\tilde{\sigma})),\nonumber
\end{eqnarray}
where:
\begin{equation}
X^1(\tilde{\tau},\tilde{\sigma})=
\frac{r_+(\tilde{\tau})\cos\phi_+(\tilde{\tau})}
{\sqrt{H^2r^2_+(\tilde{\tau})-1}}e^{-H\tilde{\sigma}},
\end{equation}
compare with Eqs.(5.20)-(5.21). After some algebra we find the integral
expression:
\begin{equation}
P_+(X^0)=-N\int d\tilde{\sigma}\frac{e^{3H(X^0-\tilde{\sigma})}
-H^2L^2e^{-H(X^0-\tilde{\sigma})}}{\sqrt{e^{4H(X^0-\tilde{\sigma})}
e^{2H(X^0-\tilde{\sigma})}+H^2L^2}},
\end{equation}
which is again of elliptic type. Asymptotically, we find the pressure
density, using also Eq.(5.25):
\begin{equation}
\frac{dP_+(X^0)}{dl_+}=-N,\;\;\;\mbox{for}\;\;X^0\rightarrow\infty
\end{equation}
Thus, in this limit the strings fulfill an equation
of state like (4.25):
\begin{equation}
dP_+(X^0)=-\frac{1}{2}dE_+(X^0),\;\;\;\mbox{for}\;\;X^0\rightarrow\infty
\end{equation}
i.e. like extremely unstable strings \cite{ven}.
\vskip 6pt
\hspace*{-6mm}{\bf II. $HL>1/2:$}\\
In this case we only find string solutions outside the horizon, as follows
by inspection of the potential, Eq.(5.1). The solutions
obtained from the stationary string ansatz are all dynamical and they are in
fact of the same type as the $r_+$ strings discussed above. We shall
therefore not go into too much detail here. We perform from the
beginning the redefinitions (5.24). Then Eq.(2.5) is solved by the
Weierstrass function (compare with Eqs.(5.5)-(5.8)):
\begin{equation}
H^2r^2(\tilde{\tau})=\frac{1}{H^2}\wp(\tilde{\tau}-
\tilde{\tau}_0;\;g_2,\;g_3)+\frac{2}{3},
\end{equation}
with invariants:
\begin{equation}
g_2=4H^4(\frac{1}{3}-H^2L^2),\;\;\;\;\;g_3=\frac{4}{3}H^6(-\frac{2}{9}+H^2L^2),
\end{equation}
discriminant:
\begin{equation}
\Delta=16H^{16}L^4(1-4H^2L^2)
\end{equation}
and roots:
\begin{equation}
e_1=\frac{H^2}{6}[-1+3i\sqrt{4H^2L^2-1}],\;\;e_2=\frac{H^2}{3},\;\;
e_3=\frac{H^2}{6}[-1-3i\sqrt{4H^2L^2-1}].
\end{equation}
The real solutions in terms of Jacobi elliptic functions take the form:
\begin{equation}
H^2r^2(\tilde{\tau})=1+HL\frac{1+\mbox{cn}[2H\sqrt{HL}\tilde{\tau},\;k]}
{1-\mbox{cn}[2H\sqrt{HL}\tilde{\tau},\;k]},
\end{equation}
where the elliptic modulus is now defined by:
\begin{equation}
k=\sqrt{\frac{1}{2}-\frac{1}{4HL}}.
\end{equation}
It follows that $k\in\;]0,\;1/\sqrt{2}[\;,$ where $k\rightarrow 0$
corresponds to $HL\rightarrow 1/2,$ while $k\rightarrow 1/\sqrt{2}$
corresponds to $HL\rightarrow\infty.$ The Azimuthal angle is
obtained by integration of Eq.(2.4). The result is most easily expressed
in terms of elliptic theta-functions:
\begin{equation}
\phi(\tilde{\tau})=\frac{1}{2i}\left\{\frac{\pi\tilde{\tau}}{\omega_2}\;
\frac{\theta_1'}{\theta_1}(\frac{\pi a}{2\omega_2})+
\log\mid\frac{\theta_1(\frac{\pi(\tilde{\tau}-a)}{2\omega_2})}
{\theta_1(\frac{\pi(\tilde{\tau}+a)}{2\omega_2})}\mid\right\},
\end{equation}
where $\omega_2=K(k)/(H\sqrt{HL})$ and $a$ is an imaginary constant
fulfilling:
\begin{equation}
a=\frac{iy}{H\sqrt{HL}},\;\;\;\;\;\mbox{cn}[2y,\;k']=\frac{1-4k^2}{3-4k^2},
\end{equation}
i.e. $y$ can be expressed as an incomplete elliptic integral. The further
analysis of the solution, Eqs.(5.47), (5.49), follows closely the analysis of
the $r_+$-solution for $HL<1/2.$ The radial coordinate, Eq.(5.47), is
periodic with period:
\begin{equation}
T_{\tilde{\tau}}=\frac{2K(k)}{H\sqrt{HL}}.
\end{equation}
For $\tilde{\tau}\in[0,\;T_{\tilde{\tau}}],$ $r$ goes from infinity towards
the horizon and back towards infinity. In the same $\tilde{\tau}$-range, the
string size which is also here in the form of Eq.(5.26),
contracts from infinite size to zero size and then expands
back to infinite size.
The Azimuthal angle increases from $\phi=0$ to $\phi=\Delta\phi:$
\begin{equation}
\Delta\phi=2\frac{\sqrt{2-4k^2}}{1-4k^2}[K(k)-\frac{2}{(3-4k^2)\sqrt{1-k^2}}
\Pi((\frac{1-4k^2}{3-4k^2})^2,\;\frac{k^2}{k^2-1})].
\end{equation}
Generally the solution describes infinitely many strings. The condition
for the solution to describe only finitely many strings is of the form (5.32),
with $\Delta\phi_+$ substituted by $\Delta\phi,$
and leads to $M/N\in\;]1/\sqrt{2}-1/2,\;1/2[.\;$ The simplest examples
are $(N,M)=(3,1),\;(N,M)=(4,1)$ etc. and the physical interpretation
of these string solutions is similar to the physical interpretation of
the $r_+$-solution discussed before for $HL<1/2.$

We close this section with a small remark on the line element (2.11) for
$a(r)=1-H^2r^2.$ It is convenient to introduce the fundamental quadratic
form $\alpha(\sigma):$
\begin{equation}
\frac{1}{2}e^\alpha=1-H^2r^2,
\end{equation}
determining the physical string size (inside the horizon). From the
$r$-equation of motion in de Sitter spacetime, Eqs.(2.8), (5.1), follows:
\begin{equation}
\frac{d^2\alpha}{d\sigma^2}+H^2e^\alpha-4H^2L^2e^{-\alpha}=0.
\end{equation}
The redefinitions $\sigma'=\sqrt{2HL}\;H\sigma,\;\;\alpha(\sigma)=
\log(2HL)+\tilde{\alpha}(\sigma')\;$ yield:
\begin{equation}
\frac{d^2\tilde{\alpha}}{d\sigma'^2}+e^{\tilde{\alpha}}-
e^{-\tilde{\alpha}}=0,
\end{equation}
that is, the sh-Gordon equation, as was proved more generally by de Vega and
S\'{a}nchez \cite{san6}.
\section{Schwarzschild Black Hole}
\setcounter{equation}{0}
Stationary strings in the background of a Schwarzschild black hole (and its
charged and rotating generalizations) were already considered in
Ref.\cite{fro}. In this section we will describe {\it all} string solutions
in the Schwarzschild black hole background obtained
from the ansatz (2.3). These will include also {\it dynamical} strings inside
the event horizon. The string potential is obtained from Eq.(2.7), for
$a(r)=1-2m/r:$
\begin{equation}
V(r)=-(1-\frac{2m}{r})[1-\frac{2m}{r}-\frac{L^2}{r^2}],
\end{equation}
see Fig.5a. The potential vanishes at the horizon $r=2m$ and for $r=r_0:$
\begin{equation}
r_0=m+\sqrt{m^2+L^2},
\end{equation}
but it is easily seen from Eqs.(2.9)-(2.10) that none of the corresponding
circular string solutions fulfill the original second order differential
equations (2.2). Thus there are no stationary circular strings in the
background of a Schwarzschild black hole.
The line element for the string solutions take the form:
\begin{equation}
ds^2=(1-\frac{2m}{r})(-d\tau^2+d\sigma^2),
\end{equation}
which defines the invariant string length element. Notice that inside the
horizon $\tau$ becomes spacelike while $\sigma$ becomes timelike, so that
the world-sheet coordinates must be redefined and the string solutions must
be expressed in better well-behaved spacetime coordinates. Only the
string
solutions outside the horizon will be stationary. Inside the horizon, the
ansatz (2.3) will describe dynamical propagating strings.

Let us first consider the solutions outside the horizon. From the potential
$V(r)$
follows that they can only exist for $r\geq r_0.$ The solutions were already
described in Ref.\cite{fro} but let us restate the results here in a somewhat
different way. Equation (2.5) is solved by the function $r_-=r_-(\sigma),$
defined by inversion of the identity:
\begin{eqnarray}
\pm\sigma\hspace{-2mm}&=&\hspace{-2mm}\frac{2m^2+L^2}{\sqrt{m^2+L^2}}
F[\psi_-(r_-),\;\frac{m^2}{m^2+L^2}]-\sqrt{m^2+L^2}\;
E[\psi_-(r_-),\;\frac{m^2}{m^2+L^2}]
\nonumber\\
\hspace{-2mm}&+&\hspace{-2mm}
m\log[\frac{2}{L^2}\sqrt{(r_-^2-2mr_--L^2)(r_-^2-2mr_-)}+
\frac{2}{L^2}(r_-^2-2mr_-)-1]\nonumber\\
\hspace{-2mm}&+&\hspace{-2mm}(r_--m)\sqrt{\frac{r_-^2-2mr_--L^2}{r_-^2-2mr_-}},
\end{eqnarray}
where:
\begin{equation}
\psi_-(r_-)=\arcsin(\sqrt{\frac{r_-^2-2mr_--L^2}{r_-^2-2mr_-}}\;)
\end{equation}
and $F$ and $E$ are the incomplete elliptic integrals of first and second
kind, respectively. The Azimuthal angle is then obtained from Eq.(2.4):
\begin{equation}
\phi_-(\sigma)=L\int_0^\sigma \frac{dx}{r_-^2(x)}=\pm\sqrt{\frac{L^2}{m^2+L^2}}
F[\psi_-(r_-(\sigma)),\;\sqrt{\frac{m^2}{m^2+L^2}}\;].
\end{equation}
Notice the following special values:
\begin{equation}
r_-(-\infty)=\infty,\;\;\;\;\;r_-(0)=r_0,\;\;\;\;\;r_-(+\infty)=\infty
\end{equation}
with corresponding angles:
\begin{equation}
\phi_-(\pm\infty)=\pm\sqrt{\frac{L^2}{m^2+L^2}}
K(\sqrt{\frac{m^2}{m^2+L^2}}\;),\;
\;\;\;\;\;\;\;\;\;\phi_-(0)=0.
\end{equation}
The solution describes an infinitely long string extending from spatial
infinity
towards the minimal distance $r_0$ from the black hole, and out towards spatial
infinity again. The angle between the two "arms" is given by:
\begin{equation}
\Delta\phi_-=2\sqrt{\frac{L^2}{m^2+L^2}}K(\sqrt{\frac{m^2}{m^2+L^2}}\;),
\end{equation}
which changes continously from $0$ to $\pi$ when $L$ changes from $0$
to $\infty.$ There are no multi-string solutions in this case; when
$\sigma$ runs from $-\infty$ to $+\infty,$ the solution describes only
one string. In the degenerate case $L=0,$ the formulas simplify
considerably. The equation (6.4) for the radial coordinate becomes:
\begin{equation}
-\sigma=r_- -4m+2m\log\frac{r_- -2m}{2m}.
\end{equation}
This string configuration
extends from spatial infinity to the horizon $r=2m,$ for
$\sigma\in]-\infty,\;+\infty]\;,$ and the angle (6.9) between the two
arms vanishes, that is, the string is straight.

Let us now consider the solutions inside the horizon. We make the
redefinitions (5.24) and introduce Kruskal coordinates:
\begin{eqnarray}
&u=\sqrt{1-\frac{r}{2m}}\;e^{r/(4m)}\sinh\frac{t}{4m},&\nonumber\\
&v=\sqrt{1-\frac{r}{2m}}\;e^{r/(4m)}\cosh\frac{t}{4m},&
\end{eqnarray}
so that (in the equatorial plane):
\begin{equation}
ds^2=\frac{32m^3}{r}e^{-r/(2m)}(-dv^2+du^2)+r^2d\phi^2.
\end{equation}
The radial coordinate
$r_+=r_+(\tilde{\tau})$ is obtained
from Eq.(2.5):
\begin{eqnarray}
\tilde{\tau}\hspace{-2mm}&=&\hspace{-2mm}\frac{2m^2+L^2}{\sqrt{m^2+L^2}}
F[\psi_+(r_+),\;\frac{m^2}{m^2+L^2}]-
\sqrt{m^2+L^2}\;E[\psi_+(r_+),\;\frac{m^2}{m^2+L^2}]
\nonumber\\
\hspace{-2mm}&-&\hspace{-2mm}m\log[\frac{-2}{L^2}
\sqrt{(r_+^2-2mr_+-L^2)(r_+^2-2mr_+)}-\frac{2}{L^2}(r_+^2-2mr_+)+1]\nonumber\\
\hspace{-2mm}&+&\hspace{-2mm}(r_+-m)\sqrt{\frac{r_+^2-2mr_+}{r_+^2-2mr_+-L^2}},
\end{eqnarray}
where:
\begin{equation}
\psi_+(r_+)=\arcsin(\sqrt{\frac{m^2+L^2}{m^2}}\;
\sqrt{\frac{r_+^2-2mr_+}{r_+^2-2mr_+-L^2}}\;)
\end{equation}
The Azimuthal angle takes the form:
\begin{equation}
\phi_+(\tilde{\tau})=L\int_0^{\tilde{\tau}}\frac{dx}{r_+^2(x)}=
\sqrt{\frac{L^2}{m^2+L^2}}
F[\psi_+(r_+(\tilde{\tau})),\;\sqrt{\frac{m^2}{m^2+L^2}}\;].
\end{equation}
Both the radial coordinate and the Azimuthal angle depend on the timelike
world-sheet coordinate $\tilde{\tau},$ only. In Kruskal coordinates
the solution is written as:
\begin{eqnarray}
u(\tilde{\tau},\tilde{\sigma})
\hspace{-2mm}&=&\hspace{-2mm}\sqrt{1-\frac{r_+(\tilde{\tau})}{2m}}
\;e^{r_+(\tilde{\tau})/(4m)}\sinh\frac{\tilde{\sigma}}{4m},\nonumber\\
v(\tilde{\tau},\tilde{\sigma})
\hspace{-2mm}&=&\hspace{-2mm}\sqrt{1-\frac{r_+(\tilde{\tau})}{2m}}
\;e^{r_+(\tilde{\tau})/(4m)}\cosh\frac{\tilde{\sigma}}{4m},\nonumber\\
\phi\hspace{-2mm}&=&\hspace{-2mm}\phi_+(\tilde{\tau}).
\end{eqnarray}
That is,
\begin{eqnarray}
v^2-u^2\hspace{-2mm}&=&\hspace{-2mm}(1-\frac{r_+(\tilde{\tau})}{2m})
e^{r_+(\tilde{\tau})/(2m)},\nonumber\\
\frac{u}{v}\hspace{-2mm}&=&\hspace{-2mm}\tanh\frac{\tilde{\sigma}}{4m},
\end{eqnarray}
and the line element becomes:
\begin{equation}
ds^2=(\frac{2m}{r_+}-1)(-d\tilde{\tau}^2+d\tilde{\sigma}^2).
\end{equation}
When $\tilde{\tau}$ goes from $\tilde{\tau}=0$ to
$\tilde{\tau}=\tilde{\tau}_0,$
where:
\begin{equation}
\tilde{\tau}_0=2\frac{2m^2+L^2}{\sqrt{m^2+L^2}}K(\frac{m^2}{m^2+L^2})-
2\sqrt{m^2+L^2}\;E(\frac{m^2}{m^2+L^2}),
\end{equation}
the radial coordinate $r_+(\tilde{\tau})$ goes from the horizon
$r_+(0)=2m$ to the spacetime singularity
$r_+(\tilde{\tau_0})=0,$ and the
solution can not be continued. In the same range of $\tilde{\tau},$ the
Azimuthal angle goes from $\phi_+(0)=0$ to:
\begin{equation}
\phi_+(\tilde{\tau}_0)=2\sqrt{\frac{L^2}{m^2+L^2}}
K(\sqrt{\frac{m^2}{m^2+L^2}}\;).
\end{equation}
It is interesting to consider also the string length element, obtained
from Eq.(6.18),
during the fall of the string towards
the singularity:
\begin{equation}
dl_+=\sqrt{2m/r_+-1}\;d\tilde{\sigma}.
\end{equation}
The physical string length is obtained by integrating this equation over
the range of $\tilde{\sigma},$ but since $r_+$ depends on $\tilde{\tau}$ only,
the string length is proportional to $\sqrt{2m/r_+-1}.$
At the horizon the string length is therefore zero, i.e. the string
starts as a point. During the fall towards the singularity, the string length
grows proportionally to $1/\sqrt{r_+}$ and eventually grows indefinetely,
see Fig.6.
This kind of behaviour for strings near the singularity of a Schwarzschild
black hole was originally found using a string series perturbation approach
\cite{lou,san2}.
Notice that the string is straight for all $\tilde{\tau}$ but because of the
$\tilde{\tau}$-dependence of the Azimuthal angle, the string falls towards
the singularity in a non-radial way.
An exceptional case is obtained for $L=0.$ In that case the Azimuthal angle
(6.14) is constant, while the radial coordinate is given by:
\begin{equation}
\tilde{\tau}=r_++2m\log\frac{2m-r_+}{2m}.
\end{equation}
The string thus falls radially from the horizon
towards the spacetime singularity
for $\tilde{\tau}\in]-\infty,\;0].$ The dynamics, for arbitrary $L,$
is in many respects
similar to what was found for the $r_+(\tilde{\tau})$-solution in de Sitter
spacetime (in the expanding phase of the de Sitter spacetime), however,
the solution in the Schwarzschild black hole background describes only
{\it one string}; no {\it multi-string}
solutions has been found in this background.
\section{2+1 Dimensional Black Hole$\;$-$\;$AdS}
\setcounter{equation}{0}
It is interesting to consider also the stationary string ansatz in the
$2+1$ dimensional black hole anti de Sitter (BH-AdS) spacetime found by
Banados et. al. \cite{ban}. A general analysis of string propagation in
the BH-AdS spacetime was performed recently by the present authors
\cite{san2}, based on the string
perturbation series approach \cite{veg2}
as well as on exact circular string configurations. Here
we will consider the new (stationary and dynamical) string solutions
(outside and
inside the horizon, respectively), obtained from the ansatz, Eq.(2.3). We will
compare these results with those obtained in Sections 4 and 6 in the equatorial
plane of ordinary anti de Sitter and Schwarzschild spacetimes.

The line element of the $2+1$ dimensional BH-AdS spacetime takes, in its
simplest version, the form:
\begin{equation}
ds^2=-(\frac{r^2}{l^2}-1)dt^2+(\frac{r^2}{l^2}-1)^{-1}dr^2+r^2d\phi^2.
\end{equation}
This line element, where $\phi$ is identified with $\phi+2\pi,$ describes
a black hole spacetime with mass $M=1$ and angular momentum $J=0.$ There is
a horizon at $r=l$ and the spacetime has constant curvature.
Asymptotically it approaches anti de Sitter spacetime
with negative cosmological constant $\Lambda=-1/l^2.$ A two-parameter
(mass $M$ and angular momentum $J$) family of black holes is obtained by
periodically identifying a linear combination of $t$ and $\phi$
\cite{ban,wel}, but the simpler
case described by Eq.(7.1) is general enough for our purposes here. Notice that
the line element (7.1) is in the general form of Eq.(2.1) with
$a(r)=r^2/l^2-1$ (and $\theta=\pi/2$). We thus obtain the string potential
Eq.(2.7):
\begin{equation}
V(r)=-(\frac{r^2}{l^2}-1)[\frac{r^2}{l^2}-1-\frac{L^2}{r^2}],
\end{equation}
see Fig.5b. The potential vanishes at the horizon $r=l$ and for $r=r_0:$
\begin{equation}
r_0=l\left( \frac{1+\sqrt{1+4L^2/l^2}}{2}\right)^{1/2},
\end{equation}
and the situation looks quite similar to the case of the Schwarzschild black
hole background. In particular, it is easily seen from Eqs.(2.9)-(2.10) that
there can be no stationary circular strings in the BH-AdS spacetime.
The line element for the string solutions reads:
\begin{equation}
ds^2=(\frac{r^2}{l^2}-1)(-d\tau^2+d\sigma^2),
\end{equation}
so that $\tau$ becomes spacelike inside the horizon while $\sigma$ becomes
timelike. As in the background of the Schwarzschild black hole, only the
strings outside the horizon will be stationary; the string solutions inside
the horizon will be dynamical.
Eq.(2.5) is solved by a Weierstrass elliptic function (compare
with Eqs.(4.4)-(4.7)):
\begin{equation}
r^2(\sigma)=l^4\wp(\sigma-\sigma_0;\;g_2,\;g_3)+\frac{2}{3}l^2,
\end{equation}
with invariants:
\begin{equation}
g_2=\frac{4}{l^6}(\frac{l^2}{3}+L^2),\;\;\;\;\;
g_3=\frac{-4}{3l^8}(\frac{2l^2}{9}+L^2),
\end{equation}
discriminant:
\begin{equation}
\Delta=\frac{16L^4}{l^{18}}(l^2+4L^2)
\end{equation}
and roots:
\begin{equation}
e_1=\frac{1}{6l^2}[-1+3\sqrt{1+4L^2/l^2}\;]\geq e_2=\frac{1}{3l^2}\geq
e_3=\frac{1}{6l^2}[-1-3\sqrt{1+4L^2/l^2}\;].
\end{equation}
We
can write Eq.(7.5) in terms of a Jacobi elliptic function:
\begin{equation}
r^2(\sigma)=l^2(\nu^2-\mu^2)+l^2\mu^2\;\mbox{ns}^2[\mu(\sigma-\sigma_0)/l,\;k],
\end{equation}
where we introduced the more compact notation (compare with Eqs.(4.8)-(4.9)):
\begin{equation}
\mu^2\equiv\sqrt{1+4L^2/l^2},\;\;\;\;\;
\nu^2\equiv\frac{1}{2}(1+\mu^2),
\;\;\;\;\;k\equiv\frac{\nu}{\mu}.
\end{equation}
It follows that the elliptic modulus $k\in\;]1/\sqrt{2},\;1],$ where
$k\rightarrow 1/\sqrt{2}$ corresponds
to $L/l\rightarrow \infty,$ while $k\rightarrow 1$ corresponds to
$L\rightarrow 0.$ We still have not fixed the integration constant $\sigma_0.$
It turns out that two qualitatively different families of real solutions appear
depending on the choice of $\sigma_0.$ For $\sigma_0=0$ we find:
\begin{equation}
r_-^2(\sigma)=l^2(\nu^2-\mu^2)+l^2\mu^2\mbox{ns}^2[\mu \sigma/l,\;k],
\end{equation}
while for $\sigma_0=\omega'=ilK'(k)/\mu:$
\begin{equation}
r_+^2(\sigma)=l^2(\nu^2-\mu^2)+l^2\nu^2\;\mbox{sn}^2[\mu \sigma/l,\;k].
\end{equation}
The solution $r_-^2(\sigma)$ is always outside the horizon and describes
stationary strings. The solution $r_+^2(\sigma),$ on the other hand, is
always inside the horizon and must be expressed in a different coordinate
system. Let us consider first the solution $r_-(\sigma)$ outside the
horizon. The corresponding Azimuthal angle is obtained from Eq.(2.4):
\begin{equation}
\phi_-(\sigma)=\frac{-k}{\sqrt{1-k^2}}\left\{ \frac{\sigma}{l}-\sqrt{2k^2-1}
\;\Pi(1-k^2,\;\mu\sigma/l,\;k)\right\}.
\end{equation}
The expressions found here for the radial coordinate and the Azimuthal angle
are formally quite
similar to the expressions found for the stationary strings in the
ordinary anti de Sitter spacetime, compare with Eqs.(4.8), (4.10), but
the elliptic modulus takes different values here and the string configurations
are actually very different.
The radial coordinate, Eq.(7.11), is periodic
with period:
\begin{equation}
T_\sigma=\frac{2lK(k)}{\mu},
\end{equation}
where $K(k)$ is the complete elliptic integral of the first kind. For
$\sigma\in[0,\;T_\sigma],\;r$ goes from infinity towards $r=r_0$ and
back towards infinity. In the same range of $\sigma,$ the Azimuthal
angle $\phi_-$ goes from $\phi_-=0$ to $\phi_-=\Delta\phi_-:$
\begin{equation}
\Delta\phi_-=2k\sqrt{\frac{2k^2-1}{1-k^2}}\;
[\Pi(1-k^2,\;k)-K(k)],
\end{equation}
compare with Eq.(4.12),
i.e. $\Delta\phi_-$ is
the angle between the two "arms" of the stationary string.
It is interesting that $\Delta\phi_-$ goes to zero in both limits
$k\rightarrow 1/\sqrt{2}$ and $k\rightarrow 1.$ In this sense the stationary
strings in the $2+1$ dimensional BH-AdS spacetime interpolates between
the stationary strings in ordinary anti de Sitter and Schwarzschild
spacetimes. The maximal angle between the arms is obtained for an
intermediate value of the elliptic modulus:
\begin{equation}
\mbox{Max}(\Delta\phi_-)=1.0023..\;\;\;\;\;\;\;\;\mbox{for}\;\;\;k=.909...
\end{equation}
As in ordinary de Sitter spacetime, the solution is actually a
{\it multi-string}
solution. In the general case the solution, Eq.(7.11), (7.13),
describes
infinitely many strings. In particular,
following the argument of Sections 4-6,
the solution describes a {\it finite}
number of strings when an equation of the
form (4.13), with $\Delta\phi_-$ given by Eq.(7.15), is fulfilled. It follows
that $M/N\in[0,\;0.1596...],$ i.e. the simplest examples are $(N,M)=(7,1),\;
(N,M)=(8,1)\;$ etc. The solution therefore describes at {\it least}
7 different strings.

We now turn to the $r_+$-solution, Eq.(7.12),
which is always inside the horizon. As in
the case of the Schwarzschild black hole (inside the horizon), we make the
redefinition Eq.(5.24) and introduce the Kruskal-like coordinates:
\begin{eqnarray}
&u=\sqrt{\frac{l-r}{l+r}}\;\sinh\frac{t}{l}&\nonumber\\
&v=\sqrt{\frac{l-r}{l+r}}\;\cosh\frac{t}{l}&
\end{eqnarray}
so that:
\begin{equation}
ds^2=(l+r)^2(-dv^2+du^2)+r^2d\phi^2.
\end{equation}
The radial coordinate now takes the form:
\begin{equation}
r_+^2(\tilde{\tau})=l^2(\nu^2-\mu^2)+
l^2\nu^2\;\mbox{sn}^2[\mu \tilde{\tau}/l,\;k],
\end{equation}
and the corresponding Azimuthal angle is obtained from
Eq.(2.4):
\begin{equation}
\phi_+(\tilde{\tau})=-k\sqrt{\frac{2k^2-1}{1-k^2}}\;\Pi(\frac{k^2}{1-k^2},
\;\mu\tilde{\tau}/l,\;k).
\end{equation}
We now follow closely the analysis of the $r_+(\tilde{\tau})$-solution inside
the horizon of the Schwarzschild black hole, see Section 6.
In Kruskal coordinates
the solution is written:
\begin{eqnarray}
u(\tilde{\tau},\tilde{\sigma})
\hspace{-2mm}&=&\hspace{-2mm}\sqrt{\frac{l-r_+(\tilde{\tau})}
{l+r_+(\tilde{\tau})}}
\;\sinh\frac{\tilde{\sigma}}{l},\nonumber\\
v(\tilde{\tau},\tilde{\sigma})
\hspace{-2mm}&=&\hspace{-2mm}\sqrt{\frac{l-r_+(\tilde{\tau})}
{l+r_+(\tilde{\tau})}}
\;\cosh\frac{\tilde{\sigma}}{l},\nonumber\\
\phi\hspace{-2mm}&=&\hspace{-2mm}\phi_+(\tilde{\tau}).
\end{eqnarray}
That is,
\begin{eqnarray}
v^2-u^2\hspace{-2mm}&=&\hspace{-2mm}\frac{l-r_+(\tilde{\tau})}
{l+r_+(\tilde{\tau})},\nonumber\\
\frac{u}{v}\hspace{-2mm}&=&\hspace{-2mm}\tanh\frac{\tilde{\sigma}}{l},
\end{eqnarray}
and the line element becomes:
\begin{equation}
ds^2=(1-\frac{r_+^2}{l^2})(-d\tilde{\tau}^2+d\tilde{\sigma}^2).
\end{equation}
When $\tilde{\tau}$ goes from $\tilde{\tau}=lK(k)/\mu\;$ to
$\tilde{\tau}=\tilde{\tau}_0,$
where:
\begin{equation}
\tilde{\tau}_0=\frac{l}{\mu}F[\arcsin(\frac{\sqrt{1-k^2}}{k}\;),\;k],
\end{equation}
the radial coordinate $r_+(\tilde{\tau})$ goes from the horizon
$r_+(lK(k)/\mu)=l$ to
$r_+(\tilde{\tau_0})=0,$ and the
solution can not be continued because of the global causal structure
of the spacetime.
In the same range of $\tilde{\tau},$ the
Azimuthal angle goes from:
\begin{equation}
\phi_+(lK(k)/\mu)=-k\sqrt{\frac{2k^2-1}{1-k^2}}\;\Pi(\frac{k^2}{1-k^2},\;k)
\end{equation}
to:
\begin{equation}
\phi_+(\tilde{\tau}_0)=-k\sqrt{\frac{2k^2-1}{1-k^2}}\;\Pi(\frac{k^2}{1-k^2},
\;\mu\tilde{\tau_0}/l,\;k).
\end{equation}
The string length element is given by:
\begin{equation}
dl_+=\sqrt{1-r_+^2/l^2}\;d\tilde{\sigma}.
\end{equation}
At the horizon the string length is zero, i.e. the string starts as a point.
When the string propagates towards $r=0,$ the string length grows but is
always {\it finite}. This illustrates the important difference between the
point $r=0$ in the Schwarzschild black hole and in the $2+1$ dimensional
BH-AdS spacetime. For the Schwarzschild black hole the point $r=0$ is
a physical curvature singularity, expressed by a power law singularity
in curvature scalars. For the $2+1$ dimensional
BH-AdS spacetime, on the other hand, the curvature is constant
$R_{\mu\nu}=-(2/l^2)g_{\mu\nu}$ everywhere, except probably at $r=0,$ where
there is at most a delta-function singularity. This difference shows up clearly
in the string solutions close to $r=0:$ In the Schwarzschild black hole
background we found that the string {\it stretches indefinetely}
near $r=0,$ which
is not the case in the $2+1$ dimensional
BH-AdS spacetime. Notice however that in {\it both}
cases the string is {\it straight}
during the non-radial fall towards $r=0.$
The infinite string stretching is a typical feature of string instability
[1,6-8,11,20] and
a generic characteristic behaviour of strings near
"strong enough" (stronger than delta-function type) spacetime singularities
\cite{med}.
\section{Schwarzschild-dS and Schwarzschild-AdS}
\setcounter{equation}{0}
Following the analysis of the preceding sections, it
is straightforward to consider also more complicated curved spacetimes from
general relativity and string theory. The mathematics will however in most
cases be quite complicated but the qualitative results can in any case
be read off directly from
the potential, Eq.(2.7). Let us illustrate this by the two examples of
ordinary Schwarzschild-anti de Sitter (S-AdS) and Schwarzschild-de
Sitter (S-dS) spacetimes.

The line element of S-AdS spacetime is given by:
\begin{equation}
ds^2=-(1-\frac{2m}{r}+H^2r^2)dt^2+(1-\frac{2m}{r}+H^2r^2)^{-1}dr^2+
r^2(d\theta^2+\sin^2\theta d\phi^2),
\end{equation}
i.e., it corresponds to $a(r)=1-2m/r+H^2r^2$ in the notation of Eq.(2.1). The
potential (2.7) takes the form:
\begin{equation}
V(r)=-(1-\frac{2m}{r}+H^2r^2)[1-\frac{2m}{r}+H^2r^2-\frac{L^2}{r^2}],
\end{equation}
see Fig.7a. The potential vanishes at the horizon $r=r_{\mbox{h}}$ and for
$r=r_0,$ where $r_{\mbox{h}}$ and $r_0$ are the {\it unique}
positive zeros of the equations:
\begin{equation}
1-\frac{2m}{r_{\mbox{h}}}+H^2r_{\mbox{h}}^2=0,\;\;\;\;\;\;
1-\frac{2m}{r_0}+H^2r_0^2-\frac{L^2}{r_0^2}=0.
\end{equation}
{}From the potential, Fig.7a., and the analysis of Sections 4,6 and 7, we can
now easily describe the qualitative features of the stationary and
dynamical string solutions obtained from the ansatz (2.3). Inside the
horizon the solution will describe one single straight
string falling non-radially
towards the physical singularity. The string starts as a point at the horizon
and stretches indefinetely for $r\rightarrow 0.$ For $r\geq r_0,$ outside the
horizon, the solution is a {\it multi-string}
solution of the same type as found
outside the horizon of the $2+1$ dimensional BH-AdS spacetime. The solution
will describe infinitely or finitely many strings depending on the value of
the parameter $L.$

In the case of S-dS spacetime, the line element is obtained from Eq.(8.1)
by changing the sign of $H^2.$ The corresponding string potential is then given
by:
\begin{equation}
V(r)=-(1-\frac{2m}{r}-H^2r^2)[1-\frac{2m}{r}-H^2r^2-\frac{L^2}{r^2}],
\end{equation}
see Fig.7b. The S-dS spacetime has two
horizons provided $\sqrt{27}Hm<1.$ We will
consider only that case. Explicit expressions for the two horizons are
given for instance in Ref.\cite{san2}. The
potential, Eq.(8.4), obviously vanishes
at the two horizons, and depending on the value of the parameter $L,$ there
can be additional (one or two more) zeros between the horizons. In Fig.7b.,
we show the most general case where the potential has four different
zeros, and all
the different types of string solutions considered until now are present.
Inside the inner (Schwarzschild-like) horizon, the string solution in S-dS is
similar to the dynamical
solution $r_+(\tilde{\tau})$
inside the horizon of Schwarzschild or Schwarzschild
anti de Sitter spacetimes, Section 6. In the
region between the two horizons a truely
stationary string of the type $r_-(\sigma)$
found inside the horizon of de Sitter spacetime
will be present, Section 5. Finally, outside
the outer (de Sitter-like) horizon the
solution describes dynamical {\it multi-strings}
of the type $r_+(\tilde{\tau})$ found outside the
horizon of ordinary de Sitter spacetime, Section 5.
\section{Conclusion}
In this paper we have studied the exact string solutions obtained by the
stationary string ansatz, Eq.(2.3),
in a variety of curved backgrounds including
Schwarzschild, de Sitter and anti de Sitter spacetimes. Many different
types of solutions have been found: closed stationary strings,
infinitely long stationary strings, dynamical straight strings and
multi-string solutions describing finitely or infinitely many stationary
or dynamical strings. In all cases we have obtained the exact solutions in
terms of either elementary or elliptic functions. Furthermore, we have
analyzed the physical properties (length, energy, pressure) of the string
solutions, thus this paper supplements earlier investigations on generic
(based on approximative methods) and exact circular string
solutions, important for the general
understanding of the string dynamics in curved spacetimes.

We close with a few remarks on the stability of the solutions. Generally, the
question of stability must be addressed by considering small perturbations
around the exact solutions. In Ref.\cite{fro2}, a covariant formalism
describing
physical perturbations propagating along an arbitrary string configuration
embedded in an arbitrary curved spacetime, was developed. The resulting
equations determining the evolution of the perturbations are however very
complicated in the general case, although partial (analytical) results
have been obtained in special cases for de Sitter
\cite{all1,all3} and Schwarzschild black hole \cite{all1,fro2,all2}
spacetimes. The exact solutions found in this paper fall
essentially into two classes: dynamical and stationary. The dynamical string
solutions outside the horizon of de Sitter (or S-dS) and inside the horizon
of Schwarzschild (or S-dS, S-AdS) spacetimes, are already unstable at the
zeroth order approximation (i.e. without including small perturbations), in
the sense that their physical length grows indefinetely. For the stationary
string solutions the situation is more delicate. The existence of the
stationary configurations is based on an exact balance between the
string tension and the local attractive or repulsive gravity. For that reason,
it can be expected that the configurations are actually unstable for certain
modes of perturbation, especially in strong curvature regions. This question
could deserve further investigations, but is out of the scope of this paper.
\setcounter{equation}{0}
\vskip 48pt
\hspace*{-6mm}{\bf Acknowledgements:}\\
A.L. Larsen is supported by the Danish Natural Science Research
Council under grant No. 11-1231-1SE.\\
The authors thank H.J. de Vega for useful discussions.

\newpage
\begin{centerline}
{\bf Figure Captions}
\end{centerline}
\vskip 24pt
\hspace*{-6mm}Fig.1. The potential $V(r),$ Eq.(2.7), in the three cases
(a) Minkowski, (b) anti de Sitter and (c) de Sitter spacetimes. The potential
is
defined such that (classical)
string solutions can only exist in the regions where $V(r)\leq 0.$
\vskip 12pt
\hspace*{-6mm}Fig.2. The $(N,M)=(5,1)$ multi-string solution in anti de
Sitter spacetime. The $(N,M)$ multi-string solutions describe $N$
stationary strings with $M$ windings in the Azimuthal angle $\phi,$
in anti de Sitter spacetime.
\vskip 12pt
\hspace*{-6mm}Fig.3. The $(N,M)=(3,2)$ stationary string solution inside the
horizon of de Sitter
spacetime. Besides the circular string, this is the simplest stationary
closed string configuration in de Sitter spacetime.
\vskip 12pt
\hspace*{-6mm}Fig.4. Schematic representation of the
time evolution of the $(N,M)=(5,1)$ dynamical multi-string solution, outside
the horizon of de Sitter spacetime. Only one of the 5 strings is shown;
the others are obtained by rotating the figure by the angles
$2\pi/5,\;4\pi/5,\;6\pi/5$ and $8\pi/5.$ During the "scattering" at the
horizon, the strings collapse to a point and re-expand.
\vskip 12pt
\hspace*{-6mm}Fig.5. The potential $V(r),$ Eq.(2.7), in the two cases
(a) Schwarzschild and (b) $2+1$ black hole anti de Sitter spacetimes.
Notice the different asymptotic behaviour in the two cases.
\vskip 12pt
\hspace*{-6mm}Fig.6. The dynamical straight string inside the horizon of
the Schwarzschild black hole. When the string falls (non-radially)
towards the singularity, the string length grows indefinetely.
\vskip 12pt
\hspace*{-6mm}Fig.7. The potential $V(r),$ Eq.(2.7), in the two cases
(a) Schwarzschild-anti de Sitter (S-AdS) and (b)
Schwarzschild-de Sitter (S-dS) spacetimes. In S-dS spacetime, degenerate
cases with less structure exist, while in S-AdS the potential takes the
form (a) for all non-zero values of the parameters $(m,H,L).$
\newpage
\begin{centerline}
{\bf Table Caption}
\end{centerline}
\vskip 24pt
\hspace*{-6mm}Table I. Short summary of the features of the string
solutions found in this paper. In anti de Sitter spacetime and outside
the horizon of de Sitter and $2+1$ BH-AdS spacetimes, the
solutions describe a {\it finite} number of strings provided a condition
of the form $N\Delta\phi=2\pi M$ is fulfilled, where $\Delta\phi$ is the
angle betwen the "arms" of the string and $(N,M)$ are integers.
\end{document}